# Current State and Challenges of Automatic Planning in Web Service Composition

**Sleiman Rabah, Dan Ni, Payam Jahanshahi, Luis Felipe Guzman**

Department of Computer Science and Software Engineering
Concordia University
Montréal, Québec, Canada
{s_rabah, d_ni, p_jahan, l_ guzman}@encs.concordia.ca

**Abstract**

This paper gives a survey on the current state of Web Service Compositions and the difficulties and solutions to automated Web Service Compositions. This first gives a definition of Web Service Composition and the motivation and goal of it. It then explores into why we need automated Web Service Compositions and formally defines the domains. Techniques and solutions are proposed by the papers we surveyed to solve the current difficulty of automated Web Service Composition. Verification and future work is discussed at the end to further extend the topic.

# 1. INTRODUCTION TO WEB SERVICE COMPOSITION (WSC)

## 1.1. Basic descriptions of Web Service and Web Service Composition

Web services are self-contained and self-describing software components that are made available and accessible across a computer network. The method in which users communicate with Web sites is through a browser that issues a HTTP request for a document given an URL, the server process the request and if the document exists and the path to it is proper then produces a response with the content of the document embedded in it; these back-and-forth messages are usually encoded in HTML. Web services are a generalization of this model so not only humans can interact with computers but also computers can communicate with other computers through the web service interfaces across the Internet, regardless of platform, programming language, etc.

The interface in which a Web service exposes its functionality is described using a neutral format called WSDL (Web Services Description Language) and it uses XML for to achieve platform/machine independence. [Figure 1] depicts the elements that compose a WSDL description. On the other hand, to implement the communication layer between Web services and potential clients it's through a standard network protocol called SOAP (Simple Object Access Protocol); also implemented using XML.

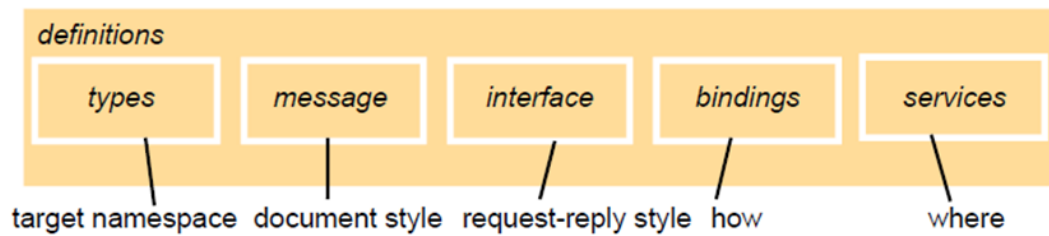

Figure 1 - Main Elements in a WSDL description

Before delving into the problem of Web service composition we need to understand how Web services are discovered and used [Figure 2]:

1. The Web service provider first publishes all the information pertinent to the Web services (WSDL, Name, Type, Location, etc.) that are offered in a discovery service.
2. A client program consults these "yellow pages" (or UDDI which stands for Universal Description Discovery and Integration) in order to locate the Web service(s) that can satisfy the request(s).

3. The client binds to the server and procures the detailed specification of the service (the WSDL file).
4. Next a SOAP message that contains the request is built and it's send to the server. In distributed system this process is known as *marshaling*.
5. The server unpacks the request (*unmarshaling*), handles it, and computes the result.
6. Lastly the response is send back in the reverse direction; *i.e.* from the server back to the client.

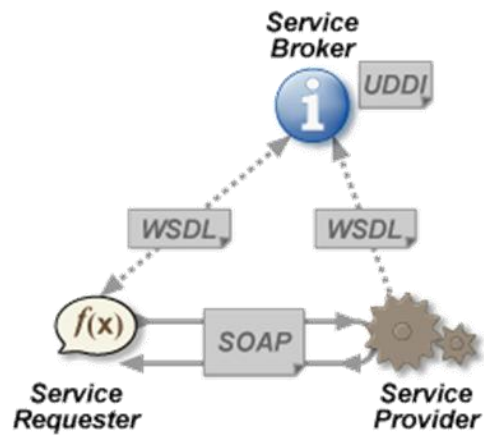

**Figure 2 - Web services architecture [Wikipedia]**

Now, the scenario described above is known as the Web service discovery (WSD) problem: Given a request *r*, among all the Web services listed in any given UDDI, how can be found a matching Web services that can satisfy *r* [1]. In this survey study, we focus on the next problem that follows WSD, that being if a single matching service does not exists that can satisfy *r*, how to compose multiple Web services that can satisfy *r*; or better known as Web service composition.

In the following subsection we explore the motivation behind WSC.

### 1.2. Motivations and Goals of Web Service Composition

"WSC can be described as a technique of composing the functionalities of relatively simpler Web services to produce a more meaningful and complex application" [8] that can generate added value to the consumer. The difficulties arise when various factors are taking into consideration; *i.e.* Web services are created and updated constantly, this propagation of WS makes "it already beyond the human ability to analysis them and generate the composition plan manually" [9]. Furthermore, rarely a request can be satisfied by a single Web service, take the motivating example of booking a tourist package: we need airplane tickets, hotel reservations, and car rental, or any combination of the three. This complexity makes necessary for this type of client requests to compose two or more services in order to produce the appropriate result.

Lastly we can mention that WSC is an iterative process; that is, the result of composing Web services is another Web service. By this property it becomes possible to use the result of WSC as a building block for further composition steps, hence more complex and valuable applications can be built (*e.g.* we can extend the tourist booking system to also include meals, attractions, discounts, deals, etc.)

We present formal models and definitions for WSC in the next section that had being defined in [1, 2].

## 2. WEB SERVICE COMPOSITION FORMAL MODELS AND DEFINITIONS

### 2.1. Definitions, Theory, Models, Formula

A Web service ($w$) can be formally described as a set of inputs $w_{in} = \{I_1, I_2, ..., I_n\}$ and a set of outputs $w_{out} = \{O_1, O_2, ..., O_n\}$. We can assume that when $w$ is invoked with all input parameters ($w_{in}$), it responds with all output parameters ($w_{out}$); in addition we have to establish that all input parameters ($w_{in}$) must be provided for $w$ to work. For a request $r$ we have a similar concept, a set of inputs ($r_{in}$) and outputs ($r_{out}$).

With this knowledge at hand, in [1] the authors formalize the WSD and WSC problems as follow:

1. **WSD** can be understood as for any request $r$ find one Web service $w$ that can solve $r$. In more formal terms: $r_{in} \supseteq w_{in}$ and $r_{out} \subseteq w_{out}$.

2. When there is no single Web service that can satisfy $r$ then there is need to compose two or more Web services in a linear and/or parallel arrangement so in conjunction a solution for $r$ can be procured. This is called **WSC**: a composition of Web services $\{w^1, w^2, ..., w^n\}$ can satisfy $r$ if first, for all $w^i \in \{w^1, w^2, ..., w^n\}$, $w^i_{in}$ can be grounded when $w^i_{out}$ is required in particular stage in the composition; and second ($r_{in} \cup w^1_{out} \cup w^2_{out} \cup ... \cup w^n_{out}) \supseteq r_{out}$.

From the above definition we can infer that in order for a WSC to satisfy the request $r$ it is necessary that for every Web service in the composition $\{w^1, w^2, ..., w^n\}$ $w^i$ can be invoked sequentially from $1$ to $n$; and that the goal state ($r_{out}$) contains the request's input parameters and all the outputs parameters for each Web service that is a member of the composition ($r_{in}$ U $w^1_{out}$ U $w^2_{out}$ U … U $w^n_{out}$).

### 2.2. Classification of the Web Service Composition problem

According to [2], the authors propose a classification schema for the Web service composition problem.

- **Manual vs. Automatic Composition**: One can do the composition manually or automatically. By doing it manually it is necessary the involvement of domain experts to analyze the problem and produce a cohesive solution, but this error-prone as per the reasons that were mentioned previously. In automatic composition software programs are involved, the most prominent being AI planning algorithms which we discuss later in this study.

- **Simple vs. Complex Operator**: Simple WSC only involves sequential composition via the AND operator "retrieve data from a web service $a_1$, AND then from $b_5$, AND then $c_9$, AND so on" [1]. Complex WSC expand this concept to include parallel processing, other operators such as OR, XOR, NOT and constraints.

- **Small vs. Large Scale**: WSC can be seen as an AI planning problem and in turn into a satisfiability problem[1]. So it is necessary to use exhaustive search algorithms (*e.g.* A*) to find a solution for small scale WSC domains. For large scale problems there is necessity for approximated algorithms though there is the risk of getting a non-optimal solution.

### 2.3. Discussions

In [2] it is described how WSC can be mapped to STRIPS (STanford Research Institute Problem Solver) planning problem: $\Pi=<P, W, r^i, r^o>$ where *P* is the set of parameters, *W* is the set of Web services in the problem domain, $r^i$ are the initial input parameters and $r^o$ are the desired output parameters. With such definition then it becomes feasible to apply AI planning methods to solve the problem of WSC. In [1] there is exposed three such planning algorithms:

---

[1] **Satisfiability** (often written in all capitals or abbreviated SAT) is the problem of determining if the variables of a given Boolean formula can be assigned in such a way as to make the formula evaluate to TRUE. This is a well know example of *NP-Complete* problems [Wikipedia].

Graphplan based planning, STATplan based reduction, and Integer Linear Programming (ILP) formulation.

These three planning methods operate in similar fashion which is: given an input in the form of a STRIPS planning problem produce, if there exists, a sequence of operations to reach the goal state. Thus, Graphplan, STATplan and ILP are good candidates when tackling the problem of automated WSC. In the following sections of this survey we delve deeper into automatic Web service composition and explore how this problem is tackled by different research articles.

## 3. AUTOMATIC PLANNING OF WEB SERVICE COMPOSITION

### 3.1. Motivation of Automated Web Services

Web service composition enables combining existing services and creating more valuable services that satisfy user requirements. It encourages reuse of existing services by discovering available interfaces; existing Web services can be reused to create new services that are able to provide more functionalities. However, as the number of available Web services proliferates, the problem of finding and combining multiple Web services takes much time and costs in the development, integration and maintenance of complex services. For this reason, automated composition becomes interesting as it can reduce costs and time in the software development cycle. Automated composition is the task of generating automatically a new Web service that achieves a given goal by interacting with a set of available Web services [5]. Therefore, automatic planning of Web services becomes a necessary next research direction in the Web services field.

### 3.2. Planning Domain and Models

By surveying on papers [4], [5], it revealed the difficulty among different planning techniques. In many realistic cases according to [5], the planner needs to deal with the nondeterministic behavior; partial observability of internal statuses, and complex goals of each individual Web services. Those are the open issues that have yet been dealt with, and were discussed in the papers [4], [5] we surveyed.

By nondeterminism [4], it means the planner cannot predict the actual interactions between external processes because it does not know if the outcome of a request turns out to be positive or negative, or if a user will accept or refuse the result a service provides. By partial obervability [4], it means the planner can only observe the communications with external processes without knowing their internal statuses or variables.

It is important to first define the planning domain, so the latter techniques or models could be further explored upon this domain. "A planning domain is defined in terms of its states, of the actions it accepts, and of the possible observations that the domain can exhibit" [4]. Some of the states are marked as valid initial states for the domain. A transition function describes how the execution of an action leads from one state to many possible different other states. Finally, an observation function defines the observation associated to each state of the domain [4].

Paper [4] formally gave three definitions of planning domain, of plan, of knowledge level domain. Those definitions serve the purpose of analyzing all planning problems in terms of mathematical models.

**Definition 1**. (Planning domain). A nondeterministic planning domain with partial observability *is a tuple*

$D = <S, A, O, I, T, X>$, *where,*

- *S is the set of* states.
- *A is the set of* actions.
- *O is the set of* observations.
- $I \subseteq S$ *is the set* of initial states; *where* $I \neq 0$.
- $T : S \times A \rightarrow 2^S$ is the transition function.
- $X : S \rightarrow O$ is the observation function.

Based on the above definition for planning domain, Definition 2 is generated to model complex plans that can encode sequential, conditional and interactive behaviors with partial observability.

**Definition 2** (Plan model). A plan for domain $D = <S, A, O, I, T, X>$ *is a tuple* $\Pi = <C, c_0, \alpha, \epsilon>$, *where:*

- *C is the set of* plan contexts.
- $c_0 \in C$ *is the* initial context.

- *α : C × O → A is the action function*

- *ϵ : C × O → C is the context evolutions function*

The contexts are the internal states of the plan. Actions to be executed, defined by function $\epsilon$, depend on the observation and on the context.

**Definition 3** (knowledge level domain). The knowledge level domain *for domain D is a tuple $D_k$ = <B, A, O, $I_B$, $T_B$, $X_B$>, where*:

- *B = {B ⊆ S: B ≠ 0 ∧ ∀ s, s'∈ B. X(s) = X(s') }.*

- *A and O are defined in domain D.*

- *$I_B$ : B x A → $2^B$ is the transition function;*

- *$X_B$ : B → O associates to each belief B the observation $X_B(B) = X(s)$ for all s ∈ B.*

To plan Web service composition, the first step is to model the protocols of external partners as planning domain as in Definition 1 & 2. With these models, we can monitor the interactions between the Web services by doing a power set of the transactions.

### 3.3 Automatic Planning Techniques

In paper [5], it proposes a technique for automated composition of Web services described in OWL-S [2] Process Model, which is to allow for the automated generation of executable processes such as BPEL4WS [3] programs. OWL-S process models [7] are declarative descriptions of the properties of Web service programs. Process models distinguish between atomic processes, i.e., non-decomposable processes that are executed by a single call and return a response, and composite processes, i.e., processes that are composed of other atomic or composite processes through the use of control constructs such as sequence, if-then-else, while loops, choice, fork, etc.

As the [Figure 3] [5] shows, Given the OWL-S process model description of n available services ($W_1,…,W_n$), each of them is encoded in a state transition system ($\sum_{W1},…, \sum_{Wn}$), and each of them describes the corresponding Web service as a state-based dynamic system, which can be partially controlled and observed by external agents [5].

---

[2] **OWL-S** is an ontology, within the OWL-based framework of the Semantic Web, for describing Semantic Web Services. It will enable users and software agents to automatically discover, invoke, compose, and monitor Web resources offering services, under specified constraints. [Wikipedia]

[3] **BPEL4WS**: Business Process Execution Language for Web Services [Wikipedia]

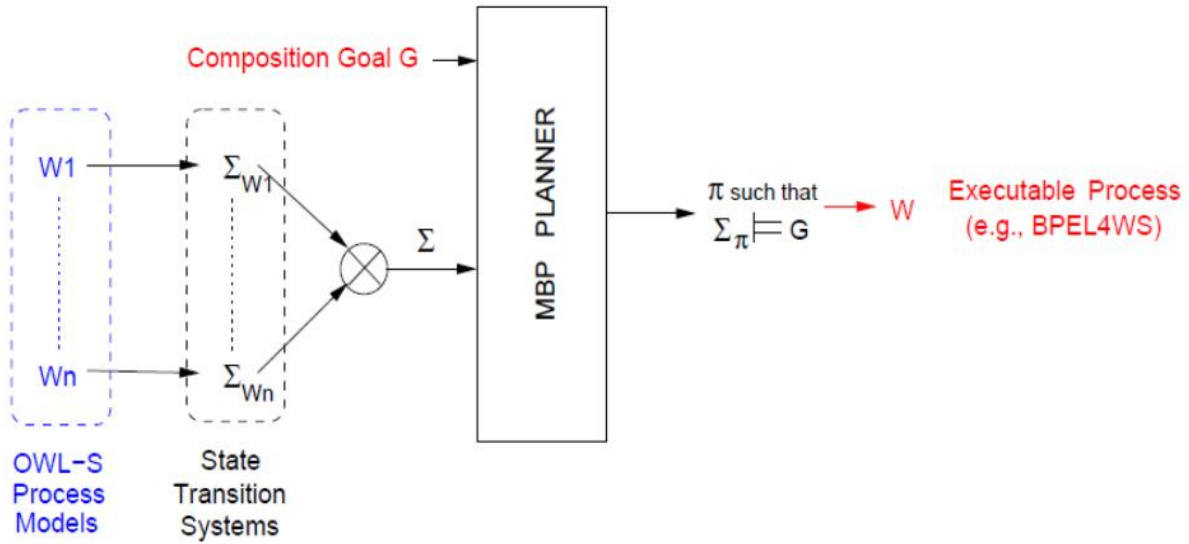

Figure 3 – OWL-S Based Automated Composition

The planning domain in this case, is a state transition system $\Sigma$ that combines $\Sigma_{W1}$,…, $\Sigma_{Wn}$. $\Sigma$ represents all the possible evolutions of the planning domain [5]. The Composition Goal G (see Figure. 2) consists of requirements on the desired behavior of the planning domain. Given $\Sigma$ and G, the planner generates a plan $\Sigma$ that interacts with the external services $W_1$,…,$W_n$ in a way such that the evolutions satisfy the goal G. The plan $\Sigma$ encodes the new service W which dynamically receives and sends invocations from/to the external services $W_1$,…,$W_n$, and which observes their behaviors, and behaves depending on responses received from the external services. The plan $\Sigma$ must therefore have the ability of encoding normal programming constructs, like tests over observations, conditionals, loops, etc. As it shows, $\Sigma$ is encoded as an automaton that, depending on the observations and on its internal state, executes different actions. $\Sigma$ can be translated into process executable languages, like BPEL4WS.

## 4. VALIDATION OF AUTOMATED WEB SERVICE COMPOSITION

Although WSC has propelled a lot of interest and research initiatives in the academia and industry areas, there exist few testing environments to evaluate these and make appropriate comparisons of performance. For automated WSC validation there are some approaches to evaluate the performance of composition and planning algorithms. Most current research in this area can be categorized in one type:

- ***Benchmark Toolkits:*** The studies in this area provide analysis of how to assess the performance of any given planning algorithm against a set of problems in the Web services discovery and composition domain. The analysis mainly focuses doing measurements on the running time it takes for a method to find a solution to a composition problem; and the number of Web services that compose the solution.

Another important field on this line of topic is Web service verification which addresses three main concerns: *information assurance*, *interoperability*, and *networthiness*. Although this area is extensive and has propelled important research initiatives, in this section of our survey study we delve into Web service composition and its validation rather than verification.

We will describe it more thoroughly in the following subsection.

### 4.1. Benchmark Toolkits

As mentioned above, one such option to evaluate WSC methods is to conduct experimentation using Benchmark testing. In [2] *Seog-Chan Oh et. al* propose an AI planning algorithm to compose Web Services (it is called Web Services Planner or WSPR); the authors validate this algorithm experimentally by using two publicly accessible test sets: EEE05 and ICEBE05. These kinds of tests are used in contests of Web Services Discovery and Composition; they contain different composition scenarios the main difference being that EEE05 contain test sets manually created by experts, and ICEBE05 are synthetically generated by semantic software (according to the authors). EEE05 is sponsored by IEEE International Conference on e-Technology, e-Commerce and e-Service and ICEBE05 by IEEE Conference on e-Business Engineering.

In general, these kinds of benchmark tools are composed of a set of automated (or manually) generated Web services files and a set of queries to test composition scenarios. More sophisticated tools are available like the one proposed in [3]. *Dongwon Lee et. al* propose a benchmark toolkit called WSBen; it offers more functionalities such as auxiliary files for statistical analysis on WSDL test sets.

## 5. CHALLENGES AND FUTURE WORK

Complex transactions that are being modeled into power set of the transaction functions are far from being optimized, and according to the experiments conducted in paper [4], it is being observed that they constitute a bottleneck. Future work would aim at a solution that avoids the computationally complex power set construction of the knowledge level domain, by providing algorithms for natively planning with extended goals under partial observability.

## 6. CONCLUSIONS

Studies in WSC are numerous and focus primarily on researching algorithms and methods for automated planning of Web services. This paper surveys state of the art WSC research in automated Web services composition planning and validation. Since WSC can be mapped to satisfiability problem, and this in turn are NP-Complete problem, there not exists a polynomial algorithm that can find the solution in a timely fashion; the time required to solve the problem using any currently known algorithm increases very quickly as the size of the problem grows. Hence, automated planning methods and approximations to find a suitable set of Web services that can solve any given request. Future work in this field can focus on lowering the complexity of WSC, or explore new approaches like D.N.A. computing or Genetic programming as means to find better algorithms to solve the WSC problem.

Different planning approaches have been proposed for the composition of Web services, but how to deal with non-determinism, partial observability, and how to generate conditional and iterative behaviors (in the style of BPEL4WS) is still an open issue.

The two papers surveyed on automated Web services compositions showed how to define the problem in the domains and models and then translate OWL-S process models to nondeterministic and partially observable state transition systems. By doing so, it automatically generates a plan that can express conditional and iterative behaviors of the composition.

# ACKNOWLEDGEMENTS

The inclusion of images and examples from external sources is only for non-commercial educational purposes, and their use is hereby acknowledged.